\documentclass{article}

\usepackage{geometry}
\usepackage{amssymb,amsbsy,amsmath,amsfonts,amscd}
\usepackage{latexsym,euscript,exscale,epic,eepic,epsfig}
\usepackage{float}
\usepackage{calc}
\usepackage{bm}
\usepackage{graphicx}
\usepackage{enumerate}
\usepackage{color}
\usepackage{pstricks,pst-node,pst-text,pst-3d,pst-plot}
\usepackage[subnum]{cases}

\usepackage{setspace}
\def \tr{\mathop{\rm tr }\nolimits}

\DeclareMathOperator*{\argmin}{arg\,min}
\def\doubleunderline#1{\underline{\underline{#1}}}

\usepackage[utf8]{inputenc}
\usepackage{verbatim}
\usepackage{listings}
\usepackage{algorithm}
\usepackage{mathtools}
\usepackage{subfig}
\usepackage{multirow}
\usepackage{enumitem}

\usepackage{authblk}
\usepackage[numbers,sort&compress,merge]{natbib}
\usepackage{hyperref}
\hypersetup{colorlinks=true, citecolor=blue}
\providecommand{\keywords}[1]{\textbf{Keywords } #1}

\title{Effects of plasticity on the anisotropy \\of the effective fracture toughness}

\author[]{Stella Brach\footnote{Corresponding author: stella.brach@polytechnique.edu}}
\affil[]{{\small LMS, CNRS, École Polytechnique, Institute Polytechnique de Paris, Palaiseau, 91128, France}}
\date{}
\begin{document}
\maketitle

\begin{abstract}
This paper investigates the effects of plasticity on the effective fracture toughness. A layered material is considered as a modelling system.
 An elastic-plastic phase-field model and a surfing boundary condition are used to study how the crack propagates throughout the material and the evolution of the effective toughness as a function of the layer angle. We first study three idealized situations, where only one property among fracture toughness, Young's modulus and yield strength is heterogeneous whereas the others are uniform. We observe that in the case of toughness and strength heterogeneity, the material exhibits anomalous isotropy: the effective toughness is equal to the largest of the point-wise values for any layer angle except when the layers are parallel to the macroscopic direction of propagation. As the layer angle decreases, the crack propagates along the brittle-to-tough interfaces, whereas it goes straight when the layers have different yield strength but uniform toughness.
We find that smooth deflections in the crack path do not induce any overall toughening and that the effective toughness is not proportional to either the cumulated fracture energy or the cumulated plastic work. In the case of elastic heterogeneity, the material is anisotropic in the sense of the effective toughness, as the latter varies as a function of the layer angle. Four toughening mechanisms are active: stress fluctuations, crack renucleation, plastic dissipation and plastic blunting. Finally, we consider a layered medium comprised of compliant-tough-weak and stiff-brittle-strong phases, as it is the case for many structural composites. We observe a transition from an interface-dominated to a plasticity-dominated failure regime, as the phase constituents become more ductile. The material is anisotropic in the sense of the effective toughness.
\end{abstract}


\section{Introduction}
Fracture toughness certainly is one of the most important property of structural materials, as it determines the lifespan of systems and devices used in applications where catastrophic failure  is unacceptable. However, despite the extremely wide impact that fracture has on both society (e.g., critical medical implants) and industry (e.g., nuclear, aerospace and civil engineering), the rational design of tough and graceful materials is still a primary challenge to modern solid mechanics. Indeed, one of the key problem is understanding how heterogeneity influences the  effective fracture properties of multi-phase materials, where phase constituents generally exhibit contrasts in elastic moduli, fracture toughness and yield strength.

There is a rich literature regarding crack propagation in multi-phase media, part of which has been driven by the desire to achieve better failure performances.
Elastically-heterogeneous materials have been subjected to extensive investigations, including  elastostatic analyses of a crack at a bi-material interface \cite{Zak-Williams-1962,Bogy-1971a,Bogy-1971b,He-1989, Cook-1972, li1997investigation}, models of elastic solids with smooth Young's modulus distributions  \cite{Dimas-2015, Dimas-2014, Gao-1991}, as well as approaches for the design of bio-inspired or 3D-printed composites with enhanced overall toughness \cite{Fratzl-2007, Murali-2011, Wang-2017}. Recent numerical and experimental studies \cite{Hossain-2014, Hsueh-2018} showed that two are the mechanisms responsible for the overall toughening of elastically-contrasted layered media: stress fluctuations induced by the elastic mismatch and crack renucleation at compliant-to-stiff interfaces.  
On the other hand, toughness heterogeneity was shown to significantly affect crack growth in particulate or fiber-reinforced composites \cite{Bower-1991,Faber-1983a, Faber-1983b, Evans-1990, Budiansky-1986, Gao-1989, Roux-2003}, hindering propagation via crack trapping and face bridging.
Crack pinning also occurs at brittle-to-tough interfaces in layered media, whose effective toughness was shown to be equal to the largest of the point-wise values \cite{Hossain-2014}.  As for the contrast in ductility, toughening due to crack trapping and bowing has been extensively reported in brittle materials reinforced with ductile particles \cite{Rubinstein-1998, Sigl-1988, Krstic-1983, Ravichandran-1992}.  Deflections in the crack path have also been observed in additively-manufactured materials \cite{zouaoui2019numerical, djouda2020experimental}, where the printing trajectory can be optimized to create textures with enhanced  toughness.
Moreover, the presence of a second-phase constituent dispersed into a matrix material can lead to orientation-dependent failure mechanisms. These have been recently investigated via different phase-field formulations, featuring orthotropic/fiber-reinforced materials with different damage/phase field variables \cite{Bleyer-2018, Lancioni-2020}.

Here, we study the relationship between heterogeneity and fracture by focusing on the effects of plasticity on the anisotropy of the effective toughness.
We refer to a layered material as a modeling system, comprised of isotropic phase constituents. 
Brach {\it et al.} \cite{Brach-JMPS-2019} studied crack propagation in layered media by using a phase-field model of brittle fracture \cite{Bourdin-2008}. Authors assumed the layer thickness to be much larger than the phase-field regularization parameter, showing that the effective toughness can depend on the overall direction of crack propagation. This work considered elastic materials with contrasts in either the elastic moduli or in the fracture toughness. The current study extends such analysis to \textit{elastic-plastic} layered materials, as plasticity is expected to greatly impact the evolution of the fracture process and thus the effective fracture toughness. To this end, we use an elastic-plastic phase-field formulation which couples elasticity, perfect plasticity and fracture in a rate-independent setting \cite{Brach-CMAME-2019,Alessi-2015}. A surfing boundary condition \cite{Hossain-2014} is applied to macroscopically propagate a crack at different layer angles.

We first consider three idealized situations, where only one property among elastic modulus, fracture toughness and yield strength is heterogeneous while the others are uniform. We show that in the case of toughness and strength heterogeneity, the material exhibits \textit{anomalous isotropy}, in the sense that the effective toughness is equal to the largest of the point-wise values for any layer angle except when the layers are disposed parallel to the macroscopic direction of  propagation. On the other hand, in presence of elastic heterogeneity, the effective toughness is \textit{anisotropic} as it varies as a function of the layer angle. Finally, we consider a layered material whose constituents are compliant-tough-weak and stiff-brittle-strong phases, as in common structural composites. We highlight the transition from an \textit{interface-dominated} to a \textit{plasticity-dominated} failure. Results show different crack propagation regimes and effective toughness depending on the layer angle. 

\textit{Notation}. Underlined and double-underlined characters are used to indicate vectors and second-order tensors respectively. Blackboard letters denote fourth-order tensors. The symbol ``$:$'' indicates the double-dot product operator. A superposed dot denotes the time derivative.

\section{Variational phase-field model}
Consider a heterogeneous body $\Omega$ comprised of linear elastic and perfectly plastic constituents. Let $\mathbb{C}$ be the stiffness tensor (with Young's modulus $E$ and Poisson's ratio $\nu$), $G_\text{c}$ the critical energy-release rate (toughness) and $\sigma_0$ the von Mises yield strength. 
We investigate the anisotropy of the effective toughness by performing variational phase-field simulations \cite{Bourdin-2000, Bourdin-2008, Francfort-1998}. As such, a regularized energy functional $\mathcal{E}_\ell$ is introduced, depending on the characteristic length $\ell>0$ and on the phase-field (or damage) variable $\alpha \in[0,1]$. The latter is equal to $0$ when the material is intact and equal to $1$ when complete fracture occurs. 
The fracture problem is solved by minimizing $\mathcal{E}_\ell$ with respect to the displacement field $\underline{u}$, the plastic strain $\doubleunderline{\epsilon}_\text{p}$ and the damage variable $\alpha$ \cite{Alessi-2015, Brach-CMAME-2019}
\begin{equation}
\label{eqn:min}
\left(\underline{u}^\ast,\doubleunderline{\epsilon}_\text{p}^\ast,\alpha^\ast\right)_\ell\,\,=\,\,\argmin_{\substack{\underline{u}\in\mathcal{K}_\text{u},\,\,\, \dot{\doubleunderline{\epsilon}}_\text{p}\in\mathcal{G},\\  \dot{\alpha}\geq 0}} \,\mathcal{E}_\ell(\underline{u},\doubleunderline{\epsilon}_\text{p},\alpha)
\end{equation}
where $\mathcal{K}_\text{u}$ is the set of kinematically-admissible displacement fields which comply with the boundary conditions and $\mathcal{G}$ is the set of plastically-incompressible strain rates (i.e., $\tr \dot{\underline{\epsilon}}_\text{p}=0$). The regularized total energy functional $\mathcal{E}_\ell$ is
\begin{equation}
\label{eqn:func}
\begin{aligned}
\mathcal{E}_\ell \big(\underline{u},\doubleunderline{\epsilon}_\text{p},\alpha\big):=&
\int_{\Omega}\frac{1}{2}\,\,(\doubleunderline{\epsilon}-\doubleunderline{\epsilon}_\text{p}):
\left(\eta+(1-\alpha)^2 \right)\mathbb{C}:(\doubleunderline{\epsilon}-\doubleunderline{\epsilon}_\text{p})\,\,d\Omega+
\int_\Omega \frac{3G_\text{c}}{8}\left[\frac{\alpha}{\ell}+\ell |\nabla\alpha |^2 \right]\,d\Omega+\\
&\int_{\Omega}\,(1-\alpha)^2\int_{0}^{\overline{t}}
\pi\big(\dot{\doubleunderline{\epsilon}}_\text{p}\big)d t\, d\Omega
\end{aligned}
\end{equation}
where $\doubleunderline{\epsilon}=(\nabla\underline{u}+\nabla\underline{u}^\text{t})/2$ is the total strain, $\eta$ is a small residual stiffness and $t$ is the time variable.  
The plastic potential is equal to $\pi\big(\dot{\doubleunderline{\epsilon}}_\text{p}\big)=\sigma_0\dot{\doubleunderline{\epsilon}}_\text{p}^\text{eq}$ if the admissibility condition $\tr \dot{\underline{\epsilon}}_\text{p}=0$ is satisfied and unbounded otherwise. The equivalent plastic strain rate is  $\dot{\epsilon}^\text{eq}_\text{p}=\sqrt{(2/3)\doubleunderline{\dot{\epsilon}}_\text{p}^\text{d}:\doubleunderline{\dot{\epsilon}}_\text{p}^\text{d}}$, with $\doubleunderline{\dot{\epsilon}}_\text{p}^\text{d}$ the deviatoric part of $\doubleunderline{\dot{\epsilon}}_\text{p}$. 

The energy functional \eqref{eqn:func} is expressed in terms of the non-dimensional quantities
\begin{equation}
    \widetilde{\mathbb{C}}=\frac{\mathbb{C}}{L_0}\,, \quad 
    \widetilde{G}_\text{c}=\frac{G_\text{c}}{E_0 L_0}\,, \quad
    \widetilde{\underline{u}}=\frac{\underline{u}}{L_0}\,, \quad
    \widetilde{\ell}=\frac{\ell}{L_0}\,, \quad
    d\widetilde{\Omega}=\frac{d\Omega}{L_0^3}
\end{equation}
where $L_0$ is a characteristic length of the domain and $E_0$ is a typical value of the Young's modulus. For the sake of compactness, the tilde appearing above the non-dimensional parameters is dropped in what follows.

In the original brittle formulation \cite{Pham-2010-I, Pham-2010-II, Marigo-2016-overview}, the evolution of the fracture process is obtained by fulfilling three principles, which are damage irreversibility (a crack can not heal), stability (the energy-release rate is bounded from above by $G_\text{c}$) and total energy balance (the crack grows when the energy-release rate is critical). In the same setting, $\Gamma$-convergence arguments \cite{Dal-Maso-2012, Braides-1998} have been used to show that the global minimizers of the regularized energy converge to those of the Griffith fracture model.
On the other hand, the asymptotic behavior of elastic-plastic functionals as in Eq.\,\eqref{eqn:func} has been studied in \cite{Dalmaso-2016}, by referring to the particular case of antiplane shear. Results showed that the $\Gamma$-limit of $\mathcal{E}_\ell$ as $\ell\to 0$ does contain a fracture energy term, which describes the crack opening.
 However, at the best of author's knowledge, the $\Gamma$-convergence of elastic-plastic functionals \eqref{eqn:func} towards a well-recognized sharp discontinuity formulation has still to be probed and represents an interesting topic of future investigation.

We note that computing the global minimizers of the energy \eqref{eqn:func} is practically not feasible, as $\mathcal{E}_\ell$ is separately convex with respect to $\alpha$ and $(\underline{u},\doubleunderline{\epsilon}_\text{p})$ but it is not jointly convex in $(\alpha, \underline{u}, \doubleunderline{\epsilon}_\text{p})$. Moreover, global minima can lead to nonphysical evolutions when the crack discontinuously propagates in time \cite{Bourdin-2011}, as it is indeed the case in heterogeneous materials. Discussions reported in \cite{Marigo-2016-overview, Tanne-2018} point to the interesting result in that evolutions which only satisfy the irreversibility and stability principles can effectively capture crack nucleation and size effects in many materials and geometries, although multiple solutions are possible. This issue has been partially addressed in \cite{Brach-CMAME-2019} by solving the fracture problem \eqref{eqn:min} via different alternating minimization algorithms, which delivered the same results. Here, the functional $\mathcal{E}_\ell$ is first minimized with respect to $(\underline{u}, \doubleunderline{\epsilon}_\text{p})$ while holding the fracture variable $\alpha$ fixed, and then minimized with respect to $\alpha$ while holding $(\underline{u}, \doubleunderline{\epsilon}_\text{p})$ fixed.

Plane strain conditions are assumed. The fracture problem is implemented using linear finite elements. We use uniform unstructured meshes with a Delaunay–Voronoi triangulation and element size $\delta$. The constrained minimization with respect to $\alpha$ is implemented by using the solvers provided by \texttt{PETSc} \cite{Balay-1997,Balay-2013a, Balay-2013b}. The minimization with respect to $\doubleunderline{\epsilon}_\text{p}$ is a constrained optimization solved with \texttt{SNLP}. The minimization with respect to $\underline{u}$ is a linear problem, implemented by using preconditioned conjugated gradients. All computations are performed via the open source code \texttt{mef90} \cite{mef90}.  The numerical fracture toughness \cite{Bourdin-2008} is $G_\text{c}^\text{num}=G_\text{c}(1+3\delta/8\ell)$, the nucleation stress \cite{Pham-2011} is $\sigma_\text{c}=\sqrt{3G_\text{c}E/(8\ell(1-\nu^2))}$ and the ductility ratio \cite{Brach-CMAME-2019} is $r_\text{y}=\sigma_\text{c}/\sigma_0$. The latter discriminates between quasi-brittle ($r_\text{y}<1$) and ductile ($r_\text{y}>1$) fracture. In the figures in the sequel, the plastic process zone is shown in green by computing the cumulated equivalent plastic strain $\epsilon_\text{p}^\text{eq}=\int_0^{\overline{t}}\dot{\epsilon}_\text{p}^\text{eq} dt$ with $\epsilon_\text{p}^\text{eq}(\overline{t})\geq 0.1\%$. Similarly, the presence of the crack is highlighted by setting $\alpha(\overline{t})\geq 0.1\%$. Finally, $\eta=10^{-6}$, $\ell=0.25$, $\delta=0.1$, $G_\text{c}^\text{num}=1.15$, $E=1$, $\nu=0.2$, $G_\text{c}=1$ and $\sigma_0=0.625$ if it is not  specified otherwise.

\begin{figure}[bt]
\centering
    {\includegraphics[width=1\textwidth]{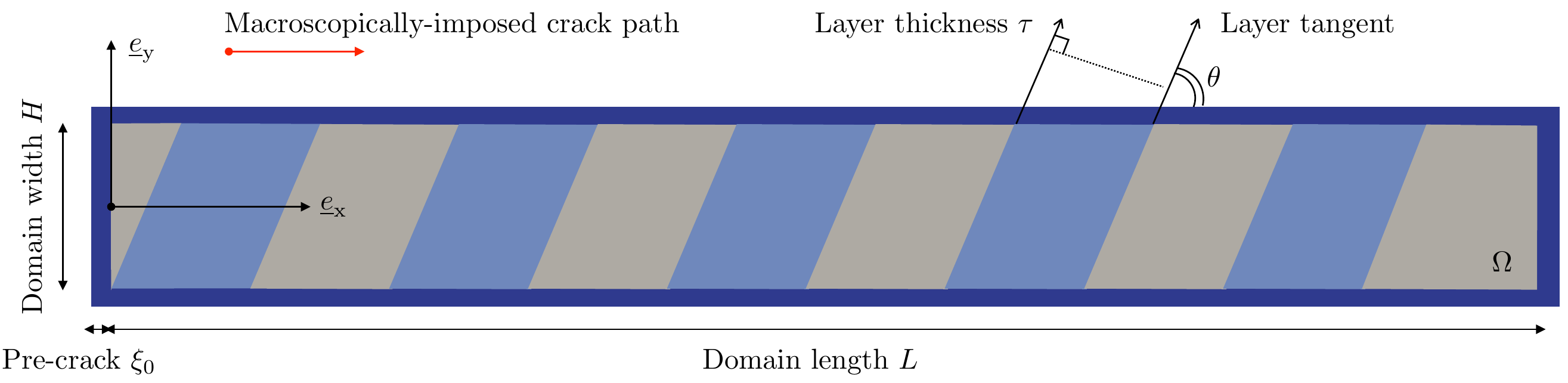}}
\caption{
{Computational domain. Notation.}}
\label{fig:domain}
\end{figure}

\subsection{Methods and prior results}\label{sec:methods}
We refer to a layered material as a modeling system. The computational domain $\Omega$ is a rectangle of length $L$ and width $H$, provided with a Cartesian reference system $(\underline{e}_\text{x}, \underline{e}_\text{y})$ (see Figure \ref{fig:domain}). The layers are oriented at an angle $\theta$ with respect to $\underline{e}_\text{x}$ and comprised of elastic-plastic materials with alternating properties. The thickness $\tau$ is defined normal to the layer axis and thus independent from $\theta$. We assume $\tau=32$, $H=40$ and $L=352$ if not otherwise specified\footnote{Recall that these are non-dimensional quantities. The results presented in this paper are independent from the dimensions of the computational domain (see also \cite{Hossain-2014, Brach-JMPS-2019}).}. Three idealized cases are firstly addressed -toughness, elastic and strength heterogeneity- where only one property among the fracture toughness $G_\text{c}$, elastic modulus $E$ and yield strength $\sigma_0$ alternates while the other two are uniform. A layered material with hybrid heterogeneity is then considered, where the layers simultaneously exhibit contrasts in strength, toughness and elastic properties. 

Following \cite{Hossain-2014}, we compute the effective fracture toughness by using a surfing boundary condition. The layered microstructure is embedded into a linear elastic and undamageable material with averaged mechanical properties. A mode-I crack opening displacement ${\underline u}({\underline{z},t}) = {\underline U}(\underline{z} - Vt \underline{e}_\text{x})$ is applied to the external boundary $\underline{z}\in\partial\Omega$ of the domain, macroscopically driving crack propagation along the horizontal direction at the velocity $V=1$. The steadily-translating displacement field reads as \cite{Zehnder-2012}
\begin{equation}
\label{eqn:surfing}
\underline{U}=\frac{\sqrt{E G_\text{c}}(1+\nu)}{E}\left(\frac{3-\nu}{1+\nu}-\cos\phi \right)\sqrt{\frac{r}{2\pi}}\left(\cos \frac{\phi}{2}\underline{e}_\text{x}+\sin \frac{\phi}{2} \underline{e}_\text{y}\right)
\end{equation}
where $(r,\phi)$ is a polar reference system emanating from the crack tip. The layers are  aligned with the macroscopic direction of propagation for $\theta=0$ and disposed perpendicular to it for $\theta=\pi/2$.  

At each time-step, we compute the effective energy-release rate as the far-field $J$-integral \cite{Rice-1968, Cherepanov-1967}
\begin{equation}
\label{eqn:Jint}
J=\int_\mathcal{L} \left(\mathcal{W}n_{\mathcal{L}_\text{x}}-\sigma_{ij}u_{i,\text{x}}n_{\mathcal{L}_j} \right)\,ds
\end{equation}
where $i,j\in\{x,y\}$, $\mathcal{W}$ is the elastic energy density, $\mathcal{L}$ is a contour embedded in the undamageable material and surrounding the crack tip, and $\underline{n}_\mathcal{L}=n_{\mathcal{L}_x}\underline{e}_\text{x}+n_{\mathcal{L}_y}\underline{e}_\text{y}$ is the outward normal to $\mathcal{L}$. The effective toughness is defined as the minimum driving force necessary to propagate a crack through macroscopic distances, that is the maximum value of the far-field $J$-integral
\begin{equation}
\label{eqn:Geff}
G_\text{c}^\text{eff}=\max_t J(t)
\end{equation}
The effective property has been shown to be independent from the temporal and spatial discretization, amplitude of the surfing loading, as well as from the particular expression used to impose the translating  opening displacement \cite{Hsueh-2018, Hossain-2014}. The anisotropy of $G_\text{c}^\text{eff}$ is investigated by propagating a crack at various angles to the layers, that is by varying the angle $\theta$.

\paragraph{Crack propagation in homogeneous elastic-plastic media}
We briefly recall here some of the results obtained in \cite{Brach-CMAME-2019}, where the variational phase-field model in Eq.\,\eqref{eqn:func} has been used to investigate crack nucleation and propagation in elastic-perfectly plastic bodies. This will be useful in what follows to understand how fracture occurs in each phase of the layered material.

\begin{figure}[hbtp]
\centering
    {\includegraphics[width=1.0\textwidth]{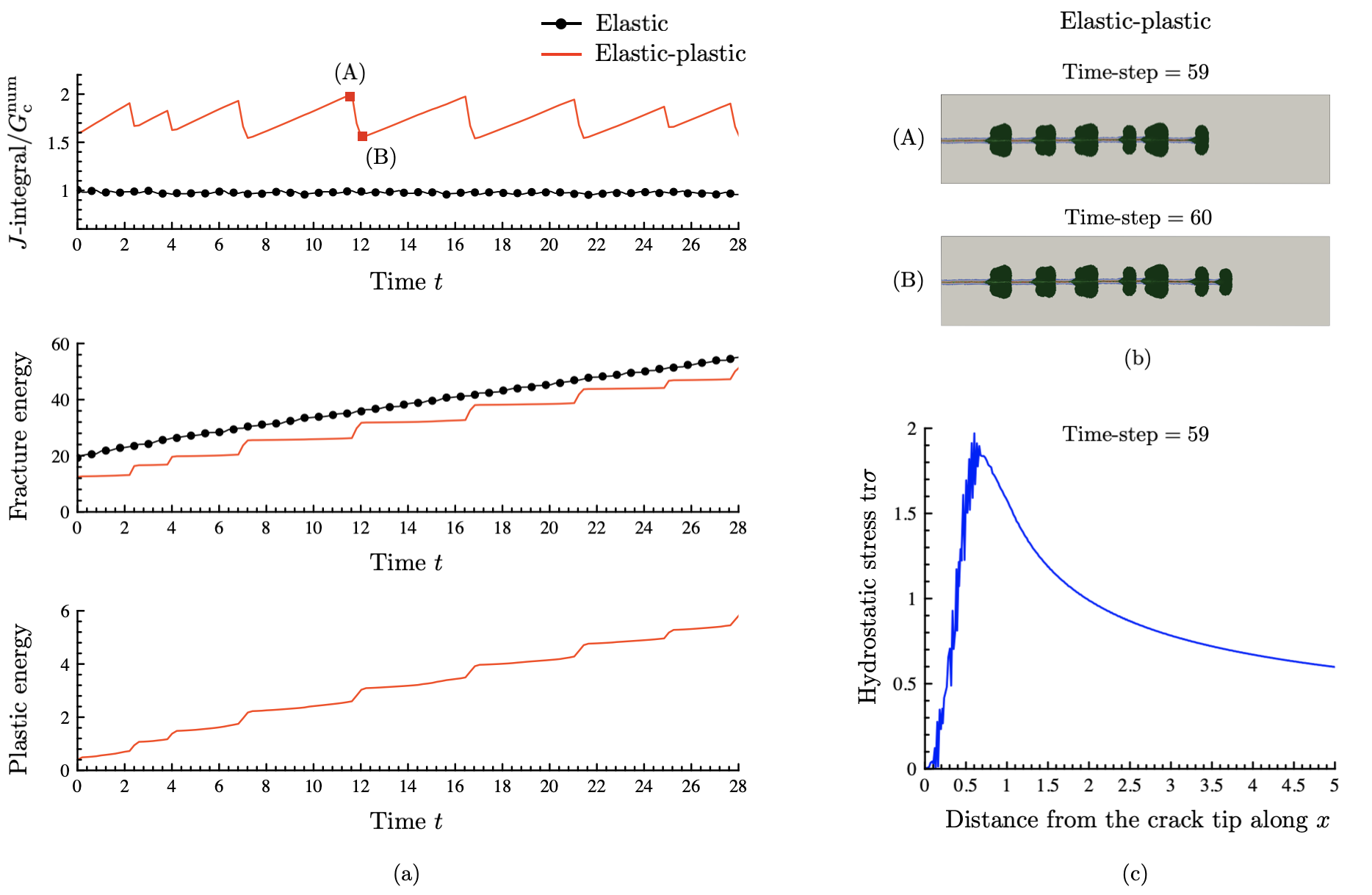}}
\caption{
{
Crack propagation in a homogeneous elastic plastic material \cite{Brach-CMAME-2019}.
(a) Normalized $J$-integral, fracture energy and cumulated plastic energy as a function of the time parameter $t$. (b) Crack propagation and plastic field distribution at two consecutive time-steps A and B. The discontinuous plastic wake shows the locations where crack propagation was arrested. (c) Evolution of the hydrostatic stress ahead of the crack tip at the time-step prior to renucleation.
}}
\label{fig:hom}
\end{figure}

Consider a homogeneous medium subjected to the surfing boundary condition \eqref{eqn:surfing}. The evolution of the $J$-integral and energies is shown in Figure \ref{fig:hom} (a) as a function of time.  In the case of an elastic material, a steady crack growth is observed as the energy-release rate equals the local toughness. On the other hand, if the material is elastic-plastic, fracture occurs in an intermittent fashion involving crack arrest and renucleation. This is shown in Figure \ref{fig:hom} (b), which displays the crack path and plastic field distribution for two consecutive time-steps A and B. Fracture proceeds as follows. 
A plastic zone develops at the crack tip hindering crack growth: the $J$-integral increases with time, while the fracture energy (which is proportional to the crack length) remains constant. 
At the point marked as A, the energy release-rate reaches a critical value for which the crack jumps forward in the material by renucleating with a plastic process zone. This corresponds to a sudden drop of the energy-release rate (point B), as well as to an increase of both fracture and cumulated plastic energies. The crack is now arrested, while the plastic zone grows under continued loading. The fracture energy thus remains constant while the plastic energy increases. Once the $J$-integral reaches a critical value, the crack jumps forward and the cycle repeats. As a consequence of this intermittent propagation, the crack leaves behind a discontinuous plastic wake, which visually marks the locations where the crack growth was arrested. Authors \cite{Brach-CMAME-2019} attributed such an intermittent propagation to the high hydrostatic stresses occurring at a certain distance ahead of the crack tip (see Figure \ref{fig:hom} (c)). In the adopted rate-independent and perfectly-plastic formulation, this triggers the crack to jump forward. Intermittent crack propagation has been experimentally reported in steel \cite{Kobayashi-1989}, and consistent with observations on crack propagation through void growth and coalescence \cite{Descatha-1981, Beremin-1981,Benzerga-2016,Pineau-2016}. Moreover, intermittent crack growth results in dimpled crack surfaces, widely documented in fractography of elastic-plastic materials (e.g., \cite{mills1987metals, lynch2006brief}).
Finally, the recent studies \cite{Dalmaso2020-jerky,Dalmaso2020-numerical}, conducted on gradient-damage models as in Eq.\,\eqref{eqn:func}, provided further proof of an intermittent crack propagation in regimes where plasticity plays a relevant role. In the particular case of an anti-plane crack, authors \cite{Dalmaso2020-jerky} demonstrated that the jerky motion first observed in \cite{Brach-CMAME-2019} is the only possible behavior.

In the present paper, the crack path and plastic field distribution computed within each layer are similar to those reported in Figure \ref{fig:hom} (b). The term ``intermittent'' is used to indicate the jerky time evolution of the fracture process, during which the propagation gets periodically arrested and the crack renucleates with a jump. The discontinuous plastic wake shown in the figures provide record of such an intermittent growth. Fracture and cumulated plastic energies are omitted for the sake of compactness, their evolution being similar to that shown in Figure \ref{fig:hom} (a).

\begin{figure}[hbtp]
\centering
\subfloat[]
    {\includegraphics[width=0.95\textwidth]{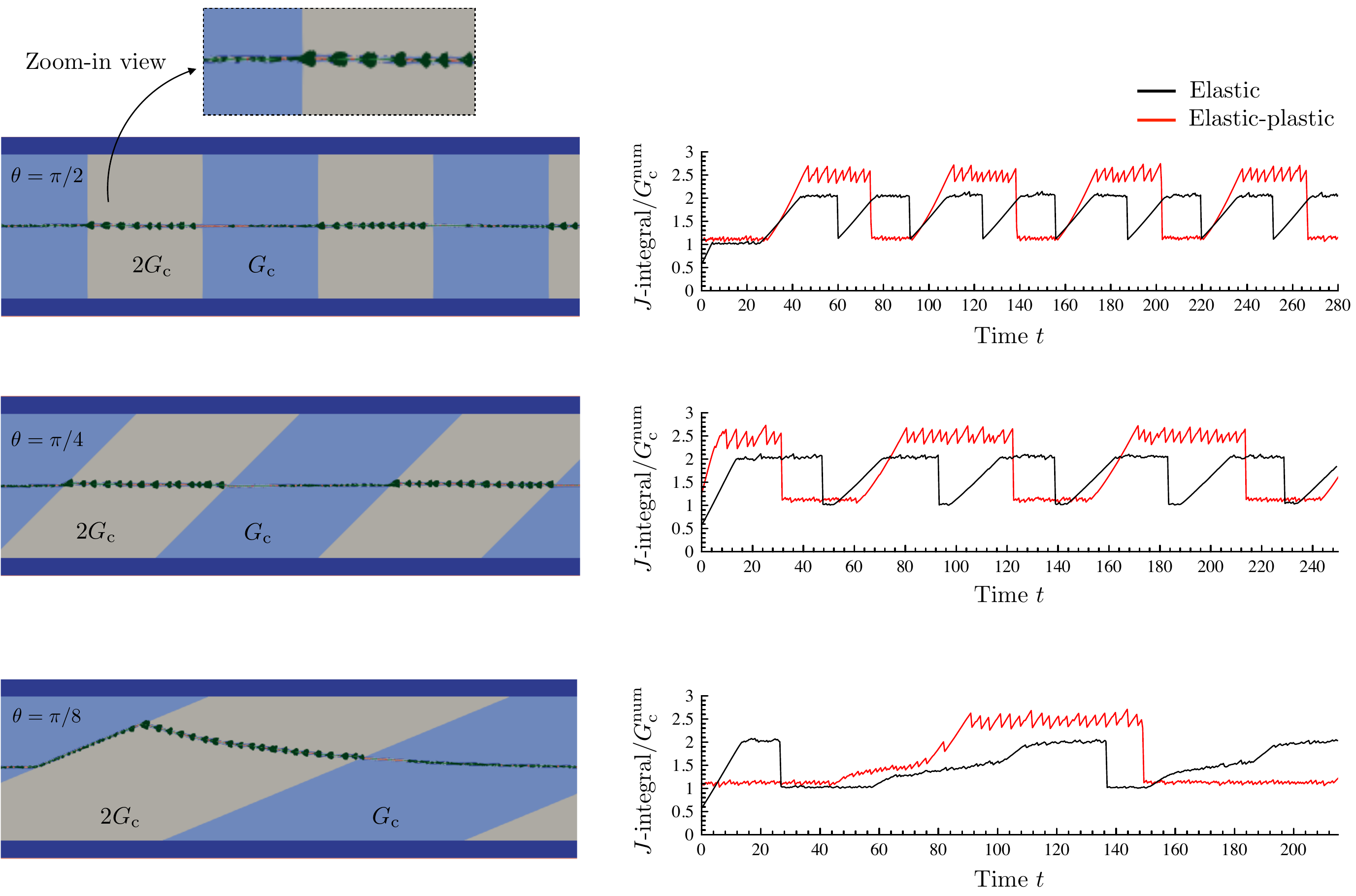}}\\
\subfloat[]
    {\includegraphics[width=0.95\textwidth]{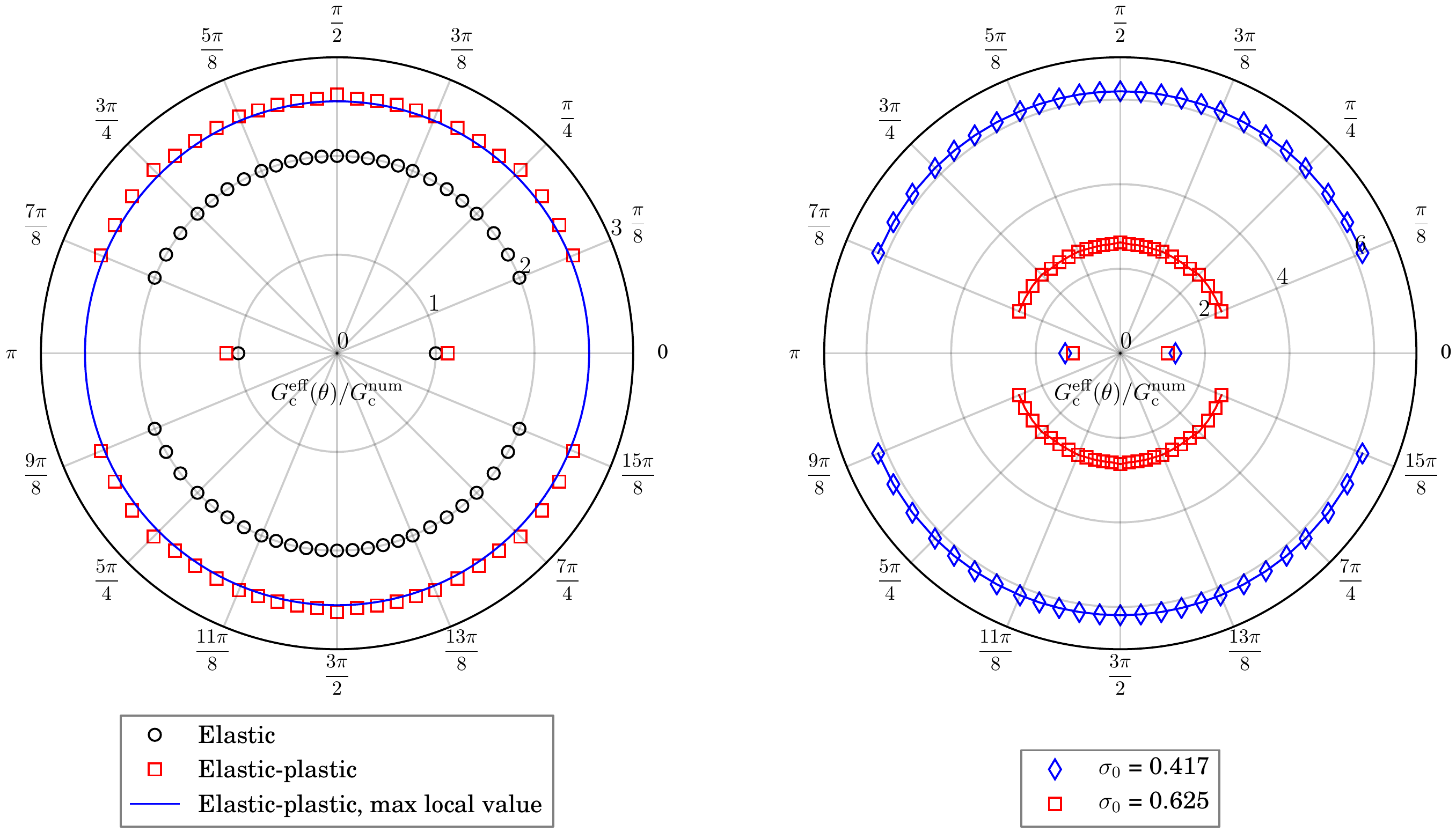}}
\caption{
{Layered material with toughness heterogeneity. (a) Crack path and far-field $J$-integral.
The discontinuous plastic wake shows the locations where crack propagation was arrested and resumed with the crack jumping forward. (b) Normalized effective toughness $G_\text{c}^\text{eff}(\theta)/G_\text{c}^\text{num}$ as a function of the layer angle $\theta$. 
In the polar plot on the left, the maximum local value identifies the toughness computed for the homogeneous elastic-plastic constituent with $G_\text{c}^2=2G_\text{c}$: the effective toughness is then equal to the largest of the point-wise properties for $\forall \theta\neq 0$, whereas it falls to the smallest of the local values for $\theta=0$. The yield strength is $\sigma_0=0.625$ if not otherwise specified.
The plastic process zone is shown in green by computing $\epsilon_\text{p}^\text{eq}(\overline{t})\geq 0.1\%$.
}}
\label{fig:tough}
\end{figure}

\section{Toughness heterogeneity}\label{sec:tough}
Consider the layered material comprised of alternating brittle (material $1$) and tough (material $2$) phases with uniform elastic moduli and yield strength. The fracture toughness is $G_\text{c}^1=G_\text{c}$ and $G_\text{c}^2=2G_\text{c}$ in material $1$ and $2$ respectively, whereas the elastic moduli are $E_1=E_2=E$, $\nu_1=\nu_2=\nu$ and the yield strength is $\sigma_0^1=\sigma_0^2=\sigma_0$. The tough material has then higher nucleation stress (i.e., $\sigma_\text{c}^2>\sigma_\text{c}^1$) and ductility ratio (i.e., $r_\text{y}^2>r_\text{y}^1$), which means that it can plastically deform more before a crack is nucleated.

Figure \ref{fig:tough}(a) shows the computed crack path and the far-field $J$-integral for various values of the layer angle $\theta$. The same figure also reports the evolution of the macroscopic energy-release rate obtained for an elastic layered medium with the same contrast in fracture toughness \cite{Brach-JMPS-2019}. 
In agreement with \cite{Brach-CMAME-2019} (see Section \ref{sec:methods}), we observe that the crack intermittently propagates within each elastic-plastic layer. In fact, a plastic process zone appears and develops at the crack tip and temporarily blunts propagation; the far-field $J$-integral correspondingly increases until reaching a critical value, which depends on the local toughness. The crack then jumps forward by nucleating a new plastic process zone; the $J$-integral drops and the cycle repeats. As a consequence of this intermittent propagation, a discontinuous plastic wake is left behind  providing record of the locations where the crack growth was arrested.  
This is more evident in the tough material, although the same observations hold for the brittle layers where plasticity concentrates in a much smaller neighborhood around the crack tip. Besides this intermittent growth within each layer, the crack is also arrested at the brittle-to-tough interfaces. As a result, the overall energy-release rate varies between the two point-wise values of toughness in the elastic case, and between the corresponding jerky evolutions in the elastic-plastic one. 

For the considered contrast, we observe that the crack grows straight throughout the microstructure for $\theta=\pi/2$, that is when the layers are perpendicular to the macroscopic direction of propagation. On the other hand, for smaller values of the layer angle $\theta$, the crack smoothly deflects at the brittle-to-tough interface and propagates along it until the energy-release rate is large enough for the crack to kick and penetrate the tough material. The resulting zig-zag path is very similar to that obtained in the elastic case \cite{Brach-JMPS-2019}. Analogous arguments allow us to conclude that the crack grows in the direction $\underline{t}$ which maximizes the energy dissipation \cite{Chambolle-2009}, that is $\underline{t}=\text{argmax}_{\underline{t}}(J(\underline{t})-G_\text{c}(\underline{t}))$.

Figure \ref{fig:tough}(b) reports the computed effective toughness $G_\text{c}^\text{eff}$ as a function of the layer angle $\theta$. 
As in the elastic case, $G_\text{c}^\text{eff}$ is equal to the maximum of the point-wise values for  $\forall\, \theta\neq 0$, whereas it falls to the smallest of the local toughnesses for $\theta=0$. In this case, the layers are parallel to the macroscopic direction of propagation and the crack only evolves through the less tough material.
Irrespective of the orientation of the microstructure, the effective toughness is larger than that computed via the elastic formulation, as a result of both plastic dissipation and plastic blunting. This becomes even more evident when considering a smaller value of yield strength $\sigma_0$, as shown in right Figure \ref{fig:tough}(b).

We thus conclude that the material exhibits \textit{anomalous isotropy}, in that the effective toughness is independent of the propagation direction for $\forall \theta \neq 0$ but it falls to the smallest of the local values for $\theta=0$.
Moreover, materials with a straight (such as for $\pi/2$) and a torturous (such as for $\pi/8$) crack path have exactly the same effective toughness. Hence, deflections in the crack path do not induce any overall toughening and the effective toughness is not proportional to the total crack length or cumulated fracture energy. By the same reasoning, this macroscopic property is not proportional to the cumulated plastic work either, as crack deflections also involve the periodic development and renucleation of plastic process zones.

\begin{figure}[hbtp]
\centering
\subfloat[]
    {\includegraphics[width=0.95\textwidth]{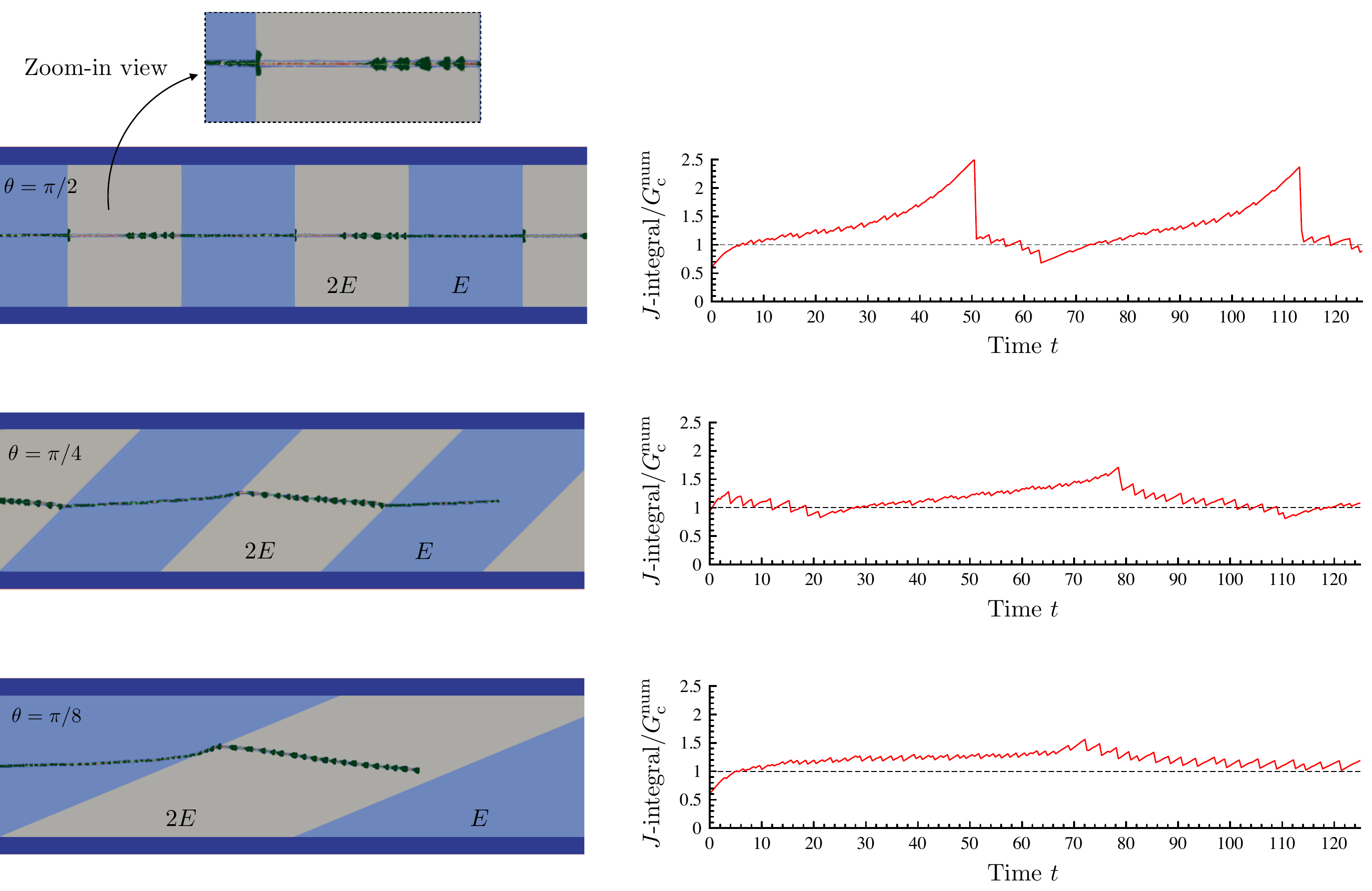}}\\
\subfloat[]
    {\includegraphics[width=0.95\textwidth]{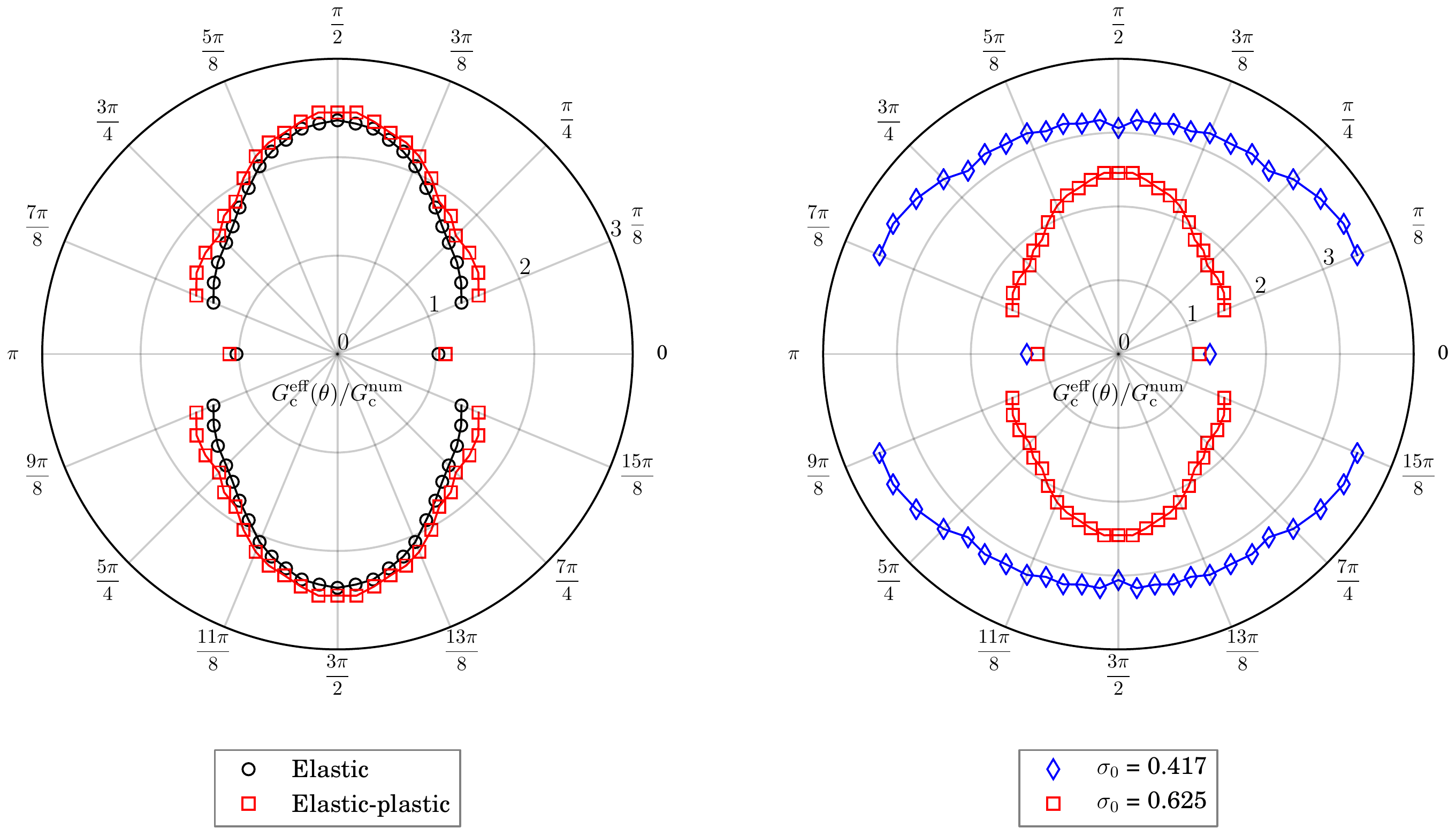}}
\caption{
{Layered material with elastic heterogeneity. (a) Crack path and far-field $J$-integral.
 The discontinuous plastic wake shows the locations where crack propagation was arrested and resumed with the crack jumping forward.  (b) Normalized effective toughness $G_\text{c}^\text{eff}(\theta)/G_\text{c}^\text{num}$ as a function of the layer angle $\theta$. The yield strength is $\sigma_0=0.625$ if not otherwise specified. The plastic process zone is shown in green by computing $\epsilon_\text{p}^\text{eq}(\overline{t})\geq 0.1\%$.}}
\label{fig:elast}
\end{figure}

\begin{figure}[h!]
\centering
    {\includegraphics[width=0.46\textwidth]{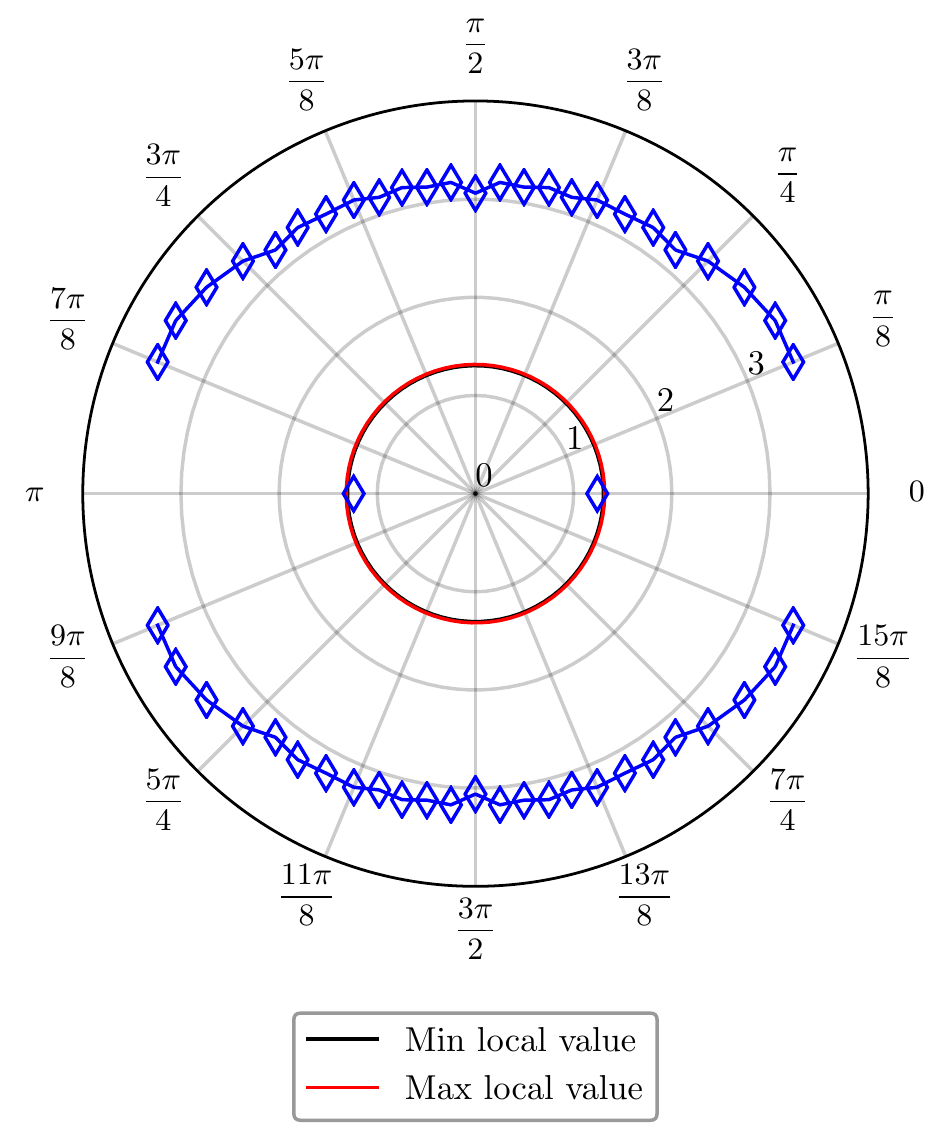}}\\
\caption{
{Layered material with elastic heterogeneity. Normalized effective toughness $G_\text{c}^\text{eff}(\theta)/G_\text{c}^\text{num}$ as a function of the layer angle $\theta$ for $\sigma_0=0.417$. The maximum (resp., minimum) local value identifies the toughness computed for the homogeneous elastic-plastic constituent with $E_2=2E$ (resp., with $E_1=E$).
The effective toughness is significantly larger than that of both elastic-plastic constituents for $\forall\theta\neq 0$, while it falls to a value close to the smallest local toughness for $\theta=0$.
}}
\label{fig:elast3}
\end{figure}

\section{Elastic heterogeneity}\label{sec:elastic}
Let the layered material be comprised of compliant (material $1$) and stiff (material $2$) phases with uniform fracture toughness and yield strength. The elastic moduli are $E_1=E$, $E_2=2E$ and $\nu_1=\nu_2=\nu$, whereas toughness and yield strength are $G_\text{c}^1=G_\text{c}^2=G_\text{c}$ and $\sigma_0^1=\sigma_0^2=\sigma_0$. The stiff material has then higher nucleation stress (i.e., $\sigma_\text{c}^2>\sigma_\text{c}^1$) and ductility ratio ($r_\text{y}^2>r_\text{y}^1$). 

The computed crack path and far-field $J$-integral are shown in Figure \ref{fig:elast} (a) for various layer angles.  The crack intermittently propagates through the compliant material by periodically renucleating with the plastic process zone. This intermittent growth is reflected by the jerky time evolution of the energy-release rate. Once reached the compliant-to-stiff interface, two regimes are observed. For $\theta=\pi/2$, the crack gets pinned at the interface and the fracture process is temporarily arrested. Under continued loading, the plastic zone expands in the stiff material inducing additional blunting. The $J$-integral increases until reaching a critical value, which is significantly higher than the uniform point-wise toughness. At this point, the crack breaks through with a jump and the $J$-integral falls to the local toughness. The propagation then resumes and the cycle repeats. 
On the other hand, as the layer angle $\theta$ decreases, the crack deviates from the macroscopic direction of propagation, advances along the interface up to a point where it gets arrested.
The $J$-integral correspondingly increases up to a critical value, which is still higher than the uniform toughness but lower than that observed for $\theta=\pi/2$. The crack then kicks and penetrates the stiff material with a jump. The $J$-integral drops and the cycle repeats.

Similar regimes of propagation have been reported in \cite{Brach-JMPS-2019} for elastic layered materials and shown to comply with the principle of local symmetry \cite{He-1989, Goldstein-1974}. Here, we observe that the distance covered by the crack along the interface is shorter than that obtained with elastic constituents, as the fracture process is considerably slowed down by plastic blunting. Moreover, a comparison among the far-field energy-release rates in Figure \ref{fig:elast}(a) shows that such  deflections in the crack path do not result in any significant toughening.

Figure \ref{fig:elast}(b) collects the values computed for the effective toughness $G_\text{c}^\text{eff}$ at various layer angles. On the left, we compare the results obtained by assuming elastic-plastic constituents with those reported in \cite{Brach-JMPS-2019} for an elastic layered medium with the same heterogeneity. We observe that the effective toughness overall decreases with decreasing the layer angle, falling to that of the compliant material when the layers are parallel to the macroscopic direction of propagation. The computed effective toughness is higher than its elastic counterpart for any angle, although the discrepancy between the two increases as $\theta$ reduces.

According to \cite{Hsueh-2018}, two are the mechanisms which are responsible of the overall toughening in elastic layered media: (i) stress fluctuations,  induced by the contrast in the elastic modulus; (ii) crack renucleation, due to the lack of driving force at compliant-to-stiff interfaces. The toughening induced by both these mechanisms decreases with decreasing the layer angle \cite{Brach-JMPS-2019}, hence explaining the evolution shown on left Figure \ref{fig:elast}(b). In elastic-plastic layered media though, two additional mechanisms occur: (iii) plastic dissipation and (iv) plastic blunting followed by crack renucleation. As shown in Figure \ref{fig:elast}(b), the contribution of these mechanisms to the overall toughening results in an opposite trend than that characterizing the elastic case, $G_\text{c}^\text{eff}$ increasing with decreasing of the layer angle. This becomes evident when considering more ductile materials, as in right Figure \ref{fig:elast}(b): the progressive reduction of the toughening coming from (i) and (ii) is compensated by the plastic contributions, leading to a material whose overall resistance to fracture is greatly enhanced at small angles.   Note also that the effective toughness of the layered material is significantly larger than that of both elastic-plastic constituents for $\forall\theta\neq 0$. This is shown in Figure \ref{fig:elast3} for the most ductile case: the normalized values of toughness of the stiff and compliant materials respectively are $1.316$ and $1.303$, while the effective property is at least three times those values for $\forall\theta\neq 0$ and close to the smallest local toughness for $\theta=0$.

We conclude that the effective toughness is \textit{anisotropic} and that plasticity tends to overall mitigate such an anisotropy, by increasing the effective toughness as the layers are oriented towards the macroscopic direction of crack propagation. On the other hand, crack deflections do not seem to significantly contribute to the overall toughening.

\begin{figure}[bt]
\centering
    {\includegraphics[width=1.0\textwidth]{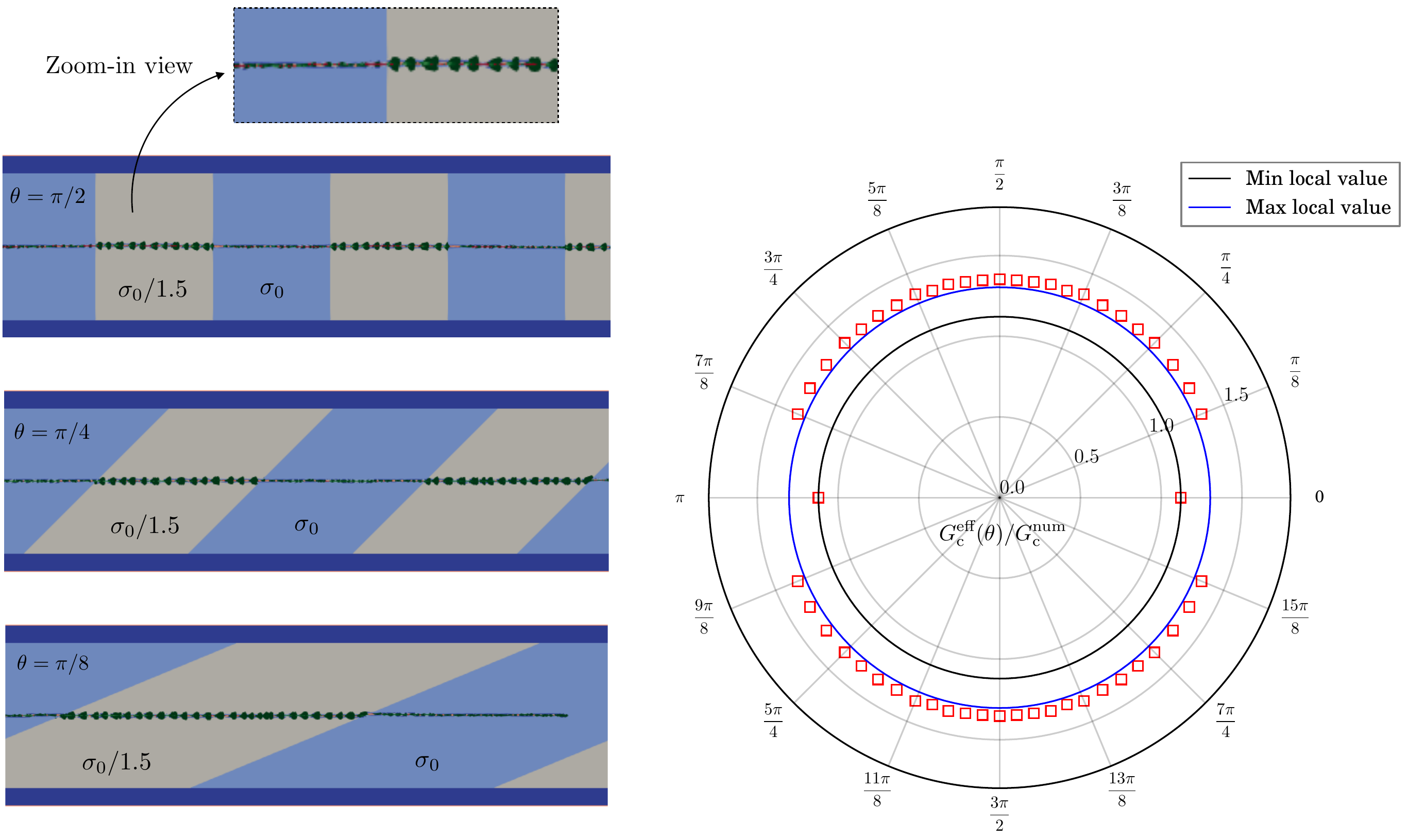}}
\caption{
{Layered material with strength heterogeneity. Crack path (left) and normalized effective toughness $G_\text{c}^\text{eff}(\theta)/G_\text{c}^\text{num}$ as a function of the layer angle $\theta$ (right). The discontinuous plastic wake shows the locations where crack propagation was arrested and resumed with the crack jumping forward. The plastic process zone is shown in green by computing $\epsilon_\text{p}^\text{eq}(\overline{t})\geq 0.1\%$.
The maximum (resp., minimum) local value identifies the toughness computed for the homogeneous elastic-plastic constituent with $\sigma_0^2=\sigma_0/1.5$  (resp., with $\sigma_0^1=\sigma_0$). The effective toughness is equal to the toughness of the weak material for $\forall\theta\neq 0$, while it falls to that of the strong constituent for $\theta=0$.}}
\label{fig:strength}
\end{figure}

\section{Strength heterogeneity}\label{sec:strength}
Consider now the alternating layers comprised of strong (material $1$) and weak (material $2$) phases with uniform elastic moduli and fracture toughness. The yield strength is $\sigma_0^1=\sigma_0$ and $\sigma_0^2=\sigma_0/1.5$, whereas the elastic moduli are $E_1=E_2=E$ and $\nu_1=\nu_2=\nu$, and the fracture toughness is $G_\text{c}^1=G_\text{c}^2=G_\text{c}$.  As such, the two constituents have equal nucleation stress $\sigma_\text{c}^1=\sigma_\text{c}^2$ and different ductility ratios $r_\text{y}^1<r_\text{y}^2$, the weak material being more ductile than the strong one.

Figure \ref{fig:strength} shows the computed crack path and the effective fracture toughness for various values of the layer angle. The evolution of the far-field $J$-integral is similar to that shown in Figure \ref{fig:tough}(a) for layered media with toughness heterogeneity and thus omitted here. We observe that the crack propagates straight throughout the microstructure, irrespective of the orientation of the layers. As fracture proceeds, the effective energy-release rate evolves between the toughness of the weak and that of the strong material for any angle except from $\theta=0$, for which the crack only evolves through the strong phase.
Hence, the effective toughness $G_\text{c}^\text{eff}$ is equal to the toughness of the weak material for $\forall\,\theta\neq0$, while it drops to that of the strong constituent for $\theta=0$.

We thus conclude that the material exhibits $\textit{anomalous isotropy}$, similarly to the case of toughness heterogeneity.

\begin{figure}[bt]
\centering
    {\includegraphics[width=0.7\textwidth]{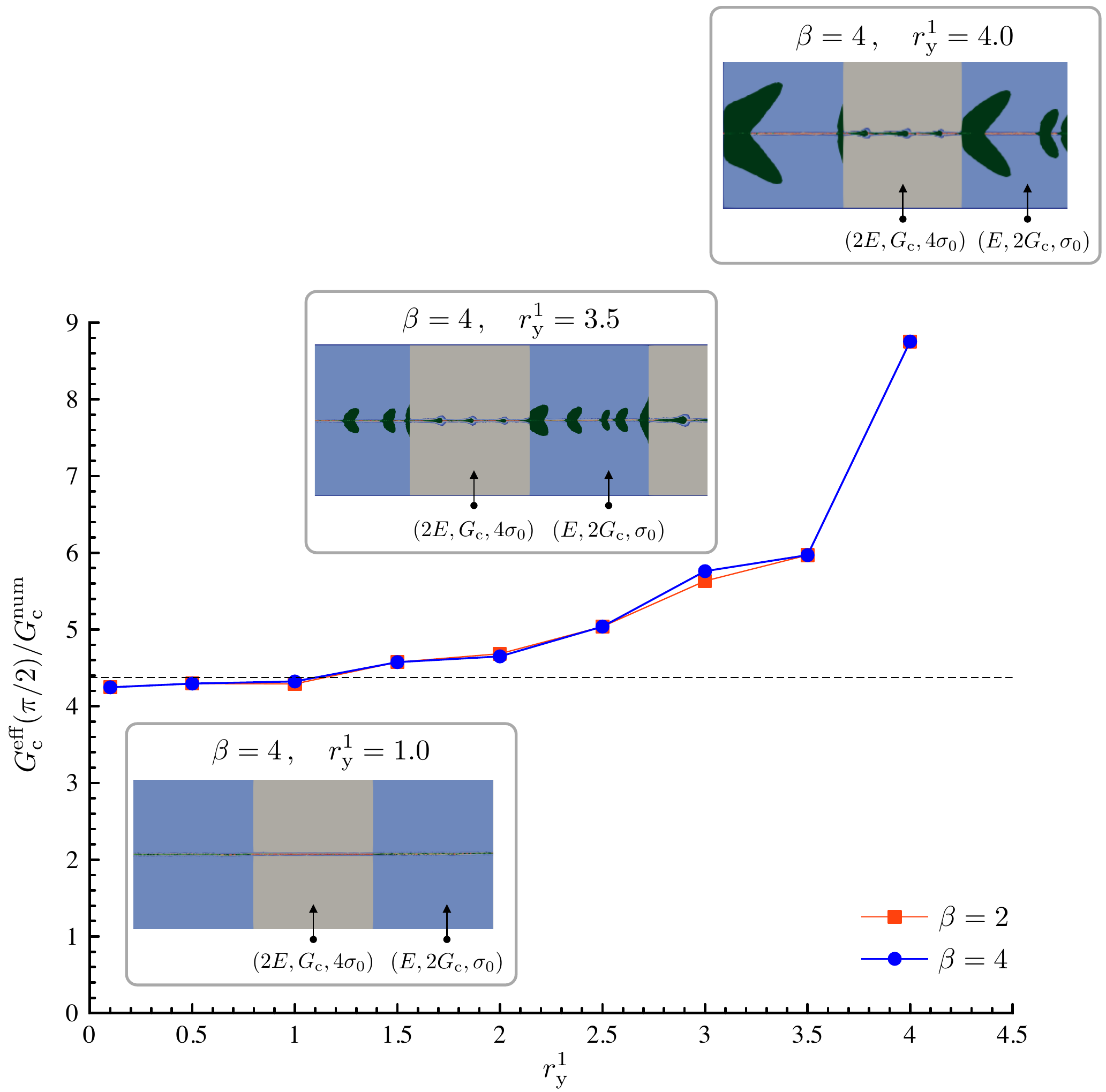}}
\caption{
{Layered material with hybrid heterogeneity. Normalized effective toughness $G_\text{c}^\text{eff}(\theta)/G_\text{c}^\text{num}$ as a function of the ductility ratio $r_\text{y}^1$ for $\theta=\pi/2$. For a given strength contrast $\beta$, the parameters $\sigma_0^1$ and $\sigma_0^2$ are computed from $r_\text{y}^1$.  The discontinuous plastic wake shows the locations where crack propagation was arrested and resumed with the crack jumping forward. The plastic process zone is shown in green by computing $\epsilon_\text{p}^\text{eq}(\overline{t})\geq 0.1\%$.}}
\label{fig:hybrid_ry}
\end{figure}

\begin{figure}[hbtp]
\centering
\subfloat[]
    {\includegraphics[width=0.95\textwidth]{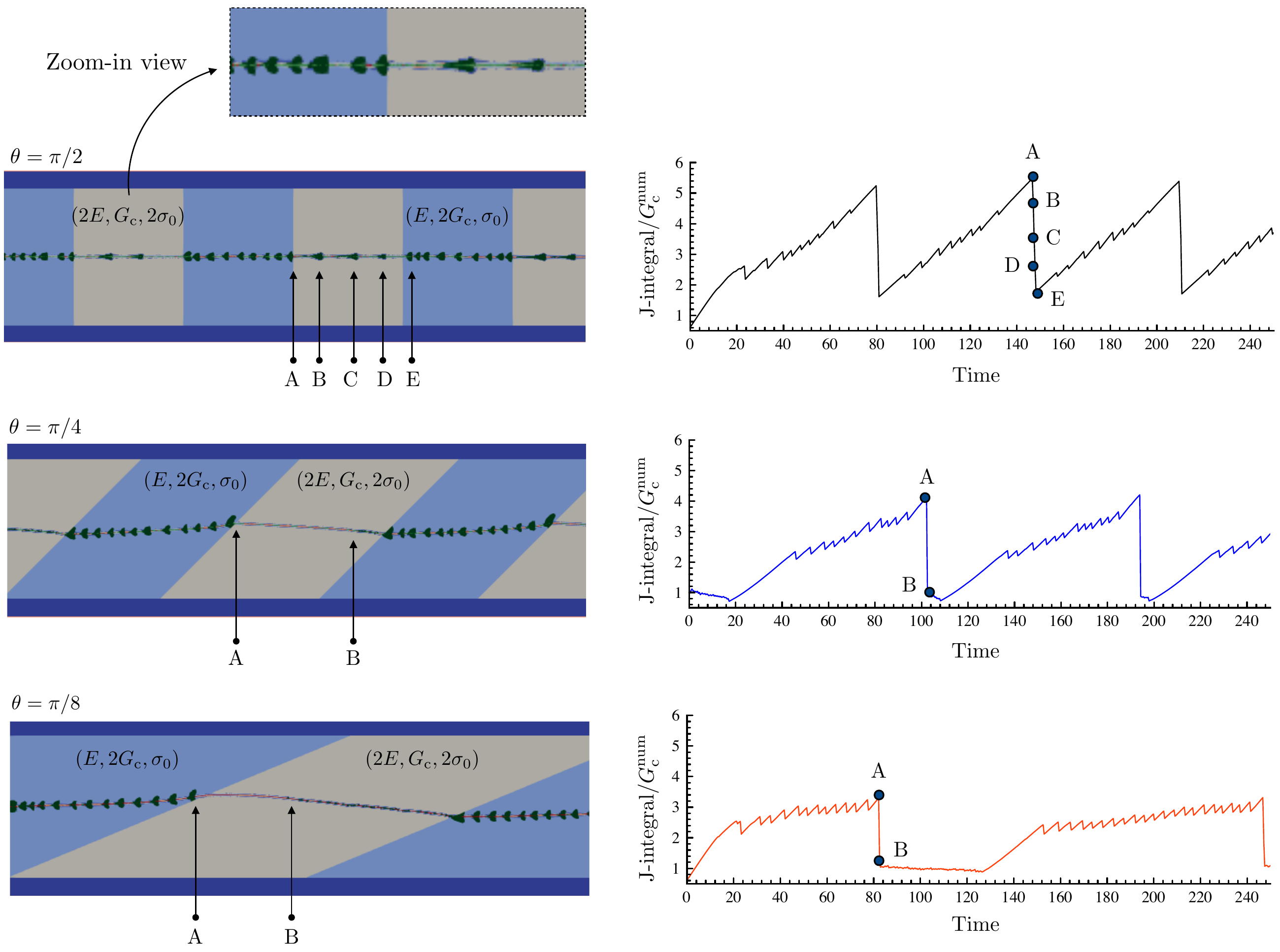}}\\
\subfloat[]
    {\includegraphics[width=0.95\textwidth]{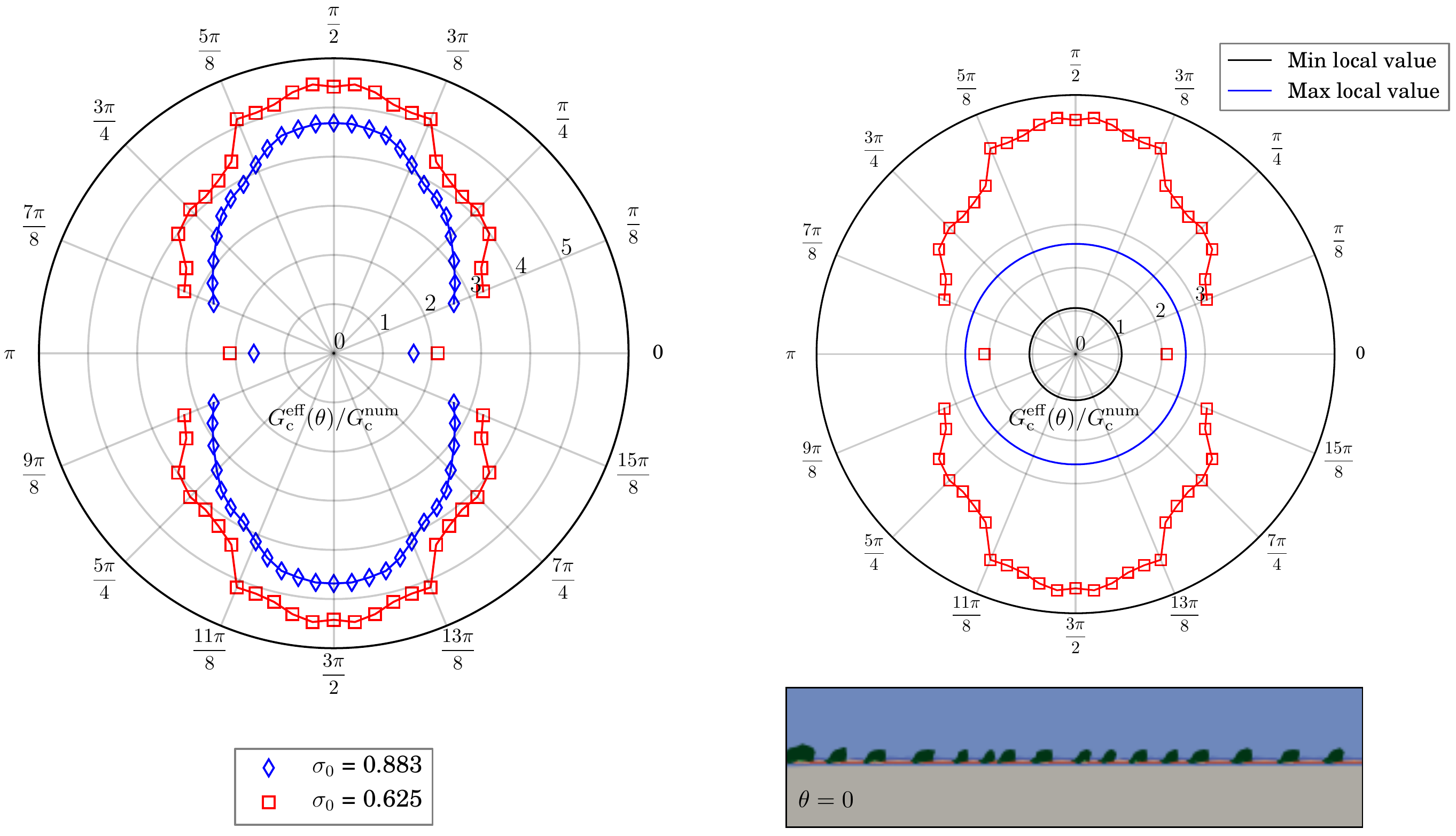}}
\caption{
{Layered material with hybrid heterogeneity. (a) Crack path and far-field $J$-integral.   (b) Normalized effective toughness $G_\text{c}^\text{eff}(\theta)/G_\text{c}^\text{num}$ as a function of the layer angle $\theta$. The yield strength is $\sigma_0=0.625$ if not otherwise specified. The plastic process zone is shown in green by computing $\epsilon_\text{p}^\text{eq}(\overline{t})\geq 0.1\%$.}
The maximum (resp., minimum) local value identifies the toughness computed for the homogeneous compliant-tough-weak (resp., stiff-brittle-strong) constituent.}
\label{fig:hybrid}
\end{figure}

\section{Hybrid heterogeneity}
While considering only one type of heterogeneity at the time is extremely helpful to understand how toughening mechanisms arise, it is to be noted that most of structural materials simultaneously exhibit contrasts in elastic modulus, toughness and yield strength. In fact, over the years the quest for better failure performances led to the development of hybrid media combining constituents with mutually exclusive properties. As such, a tough but weak phase (ensuring resistance to fracture) was often combined to a strong but brittle one (providing hardness). Examples include for instance ceramic/metal and ceramic/polymer composites \cite{Bonderer-2008, Munch-2008, Barthelat-2010}.

Here, we investigate fracture in such multi-phase media by considering a layered material comprised of compliant-tough-weak (material $1$) and stiff-brittle-strong (material $2$) phases. The elastic moduli are $E_1=E$, $E_2=2E$ and $\nu_1=\nu_2=\nu$, the fracture toughness is $G_\text{c}^1=2G_\text{c}$ and $G_\text{c}^2=G_\text{c}$, whereas the yield strength is $\sigma_0^1=\sigma_0$ and $\sigma_0^2=\beta\sigma_0$ with $\beta>1$. The constituents have then equal nucleation stress (i.e., $\sigma_\text{c}^1=\sigma_\text{c}^2$) but different ductility ratio ($r_\text{y}^2=r_\text{y}^1/\beta$). The parameter $\beta$ allows to vary the contrast in ductility between the two materials, the brittle-strong phase becoming less ductile as $\beta$ increases.

First, consider the layers to be perpendicular to the macroscopic direction of propagation. Figure \ref{fig:hybrid_ry} shows the evolution of the effective toughness as a function of $r_\text{y}^1$ for two values of the strength contrast $\beta$. The dashed line indicates the sum of the values $G_\text{c}^\text{eff}(\pi/2)/G_\text{c}^\text{num}$ obtained in the absence of plasticity and by separately considering elastic and toughness heterogeneity. 
We observe a transition between two failure regimes. 

For $r_\text{y}^1<1$, both materials are quasi-brittle and the effective toughness is independent from the ductility ratio. Since no additional toughening comes from the heterogeneity in the local toughness (see Section \ref{sec:tough}), the observed enhancement in $G_\text{c}^\text{eff}$ is due to the elastic contrast that the crack has to face when evolving from the compliant-tough to the stiff-brittle material. 
We refer to this regime as \textit{interface dominated}.
On the other hand, for $r_\text{y}^1>1$, the constituents are more ductile and the effective toughness increases with increasing $r_\text{y}^1$. For high ductility ratios, a plastic zone consisting of two lobes appears at the tip of the crack located in the tough material, by temporarily arresting propagation. Then, the crack directly jumps forward at the compliant-to-stiff interface, where it stops again due to the unfavorable contrast in elastic modulus. The plastic zone grows on the compliant side of the interface causing additional blunting. The computed value of effective toughness is much larger than those obtained for $r_\text{y}^1<1$, as the fracture process is now essentially driven by plastic blunting. We thus refer to this regime as \textit{plasticity dominated}. Results in Figure \ref{fig:hybrid_ry} are practically independent from the relative ductility $\beta$ of the two materials. 

We now set $\beta=2$ and investigate how the fracture process evolves when varying the layer angle $\theta$. The ductility ratio is $r_\text{y}^1\approx 3$ (i.e., $\sigma_0=0.625$), aiming to capture features of both the interface-dominated and plasticity-dominated regimes.

Figure \ref{fig:hybrid}(a) shows the computed crack path and $J$-integral at various layer angles. The crack intermittently propagates through the tough material towards the compliant-to-stiff interface. The contrast in toughness with the upcoming layer does not affect the evolution of the $J$-integral (with respect to what observed in Section \ref{sec:tough}), which instead significantly increases as a result of the upcoming mismatch in elastic modulus. Once reached the compliant-to-stiff interface, the crack is arrested and the plastic zone entirely develops within the compliant-tough material. The $J$-integral then increases up to a critical value for which the propagation becomes unstable and the crack jumps forward. At this point, we observe two different regimes of propagation depending on the layer angle. 

For $\theta=\pi/2$, the crack goes straight through the stiff material by a rapid sequence of jumps the last of which is within the compliant-tough layer. The $J$-integral drops almost instantaneously to the local value of toughness and starts increasing again due to the upcoming elastic contrast. 
On the other hand, for smaller values of $\theta$, the crack gets deflected from the macroscopic direction of propagation. However, differently from what observed in Section \ref{sec:elastic}, it does not propagate along the interface. In fact, the contrast in ductility between the two layers is such that the plastic zone preferably develops in the tough material. This prevents the crack to advance any further along the interface, finding more convenient to renucleate in the brittle material despite the higher elastic modulus. 
The propagation is then driven by the surfing condition back to the macroscopic path and the crack steadily advances as the $J$-integral equals the local toughness (see $\theta=\pi/8$).

Figure \ref{fig:hybrid}(b) shows the effective toughness at various layer angles, for two values of yield strength. Similar to the case of elastic heterogeneity, $G_\text{c}^\text{eff}$ decreases with decreasing $\theta$, attaining values strictly larger than the toughness of the constituents for $\forall\,\theta\neq0$. At $\theta=0$, the crack propagates at the compliant interface, however the plastic process zone only develops within the compliant material. Because of that, $G_\text{c}^\text{eff}$ drops to a value which is smaller than the toughness of the compliant-tough material. Interestingly, for a low yield strength and more ductile layers,  the effective toughness does not varies smoothly as a function of $\theta$, as it is shown to be the case for $r_\text{y}^1=2$ (i.e., $\sigma_0=0.883$). This is due to the abrupt transition between the first (crack pinned perpendicularly at the interface) and the second (crack deflection and jump) regime of propagation.

The material is \textit{anisotropic} in the effective toughness, similarly to the case of elastic heterogeneity.

\section{Conclusions}
This paper investigates the anisotropy of the effective fracture toughness \cite{Brach-JMPS-2019}, with a focus on the effects induced by plasticity. We consider elastic-plastic layered materials with various contrasts in the local properties. The effective  toughness is determined by combining the elastic-plastic phase-field model \cite{Bourdin-2008,Brach-CMAME-2019} with the surfing boundary condition \cite{Hossain-2014}.
We study how the crack propagates throughout the material and the evolution of the effective toughness as a function of the layer angle.

We start by considering three idealized situations where only one property among fracture toughness, elastic modulus and yield strength varies while the others are uniform. 

In the case of toughness and strength heterogeneity, the layered material exhibits \textit{anomalous isotropy} as the effective toughness is equal to the largest of the point-wise values for any layer angle except when the layers are disposed parallel to the macroscopic direction of propagation. In this case, the effective toughness is equal to the smallest of the local toughnesses. As the layer angle decreases, the crack gets deflected at the brittle-to-tough interfaces whereas it propagates straight when the layers have uniform toughness but alternating  strength.
We observe that smooth deflections in the crack path do not induce any significant overall toughening and that the effective toughness is not proportional to either the cumulated fracture energy or the cumulated plastic work.

In the case of elastic heterogeneity, the layered material is \textit{anisotropic} in the sense of the effective toughness, as the macroscopic property decreases with decreasing the layer angle. An important observation here is that the effective toughness is strictly larger than the uniform value as long as the crack interacts with both the constituents.
According to \cite{Hsueh-2018}, in the absence of plasticity two are the mechanisms which are responsible for such an enhancement of toughness: stress fluctuations and crack renucleation at compliant-to-stiff interfaces. Our computations highlight the occurrence of two additional contributions: plastic dissipation and plastic blunting followed by crack renucleation. These mechanisms become dominant when considering more ductile constituents, leading to an increase in the effective toughness as the layer angle reduces. 

Finally, we consider layers comprised of alternating compliant-tough-weak and stiff-brittle-strong phases, as it is the case in many composites relevant for engineering applications. We perform two analyses. 
In the first one, we fix the layer angle and study how the effective toughness varies when considering more ductile constituents. We observe a transition from an \textit{interface-dominated} to a \textit{plasticity-dominated} regime. In the first situation, the overall toughening is mainly due to stress fluctuations and crack renucleation whereas in the second scenario plastic dissipation and blunting become the dominant toughening mechanisms. The obtained results do not significantly depend on the mismatch in yield strength. This suggests that contrasts in elastic modulus and fracture toughness, rather than strength heterogeneity, are the major players involved in the fracture process of such composites. 
In the second analysis, we fix the ductility of the constituents and study how the fracture evolves as a function of the layer angle. We observe two regimes of propagation. In the first, the crack goes straight throughout the microstructure and interfaces are plastically deformed. In the second, the crack is deflected from the macroscopic path while approaching the compliant-to-stiff interface; here, the plastic zone prevents the crack to further grow along the interface, thus triggering instability. The layered material is \textit{anisotropic} in the sense of the effective toughness.

\section*{Acknowledgement}
I am delighted to acknowledge Profs. Kaushik Bhattacharya and Blaise Bourdin for their guidance and advices.
I gratefully acknowledge the financial support of the U.S. National Science Foundation (Grant No. DMS-1535083 and 1535076) under the Designing Materials to Revolutionize and Engineer our Future (DMREF) Program. The development of the numerical codes used as part of this project was supported in part by a grant from the National Science Foundation DMS-1716763. The numerical simulations were performed at the Caltech high performance cluster supported in part by the Moore Foundation.

\bibliography{mainb}

\end{document}